\documentclass[10pt,conference]{IEEEtran}
\IEEEoverridecommandlockouts

\usepackage{cite}
\usepackage{amsmath,amssymb,amsfonts}
\usepackage{graphicx}
\usepackage{textcomp}
\usepackage{xcolor}

\usepackage{multirow}
\usepackage{tabularx}
\usepackage{adjustbox}
\usepackage{booktabs}
\usepackage{caption}
\usepackage{enumitem}
\usepackage{array}
\usepackage{wrapfig}
\usepackage{amsmath,amsfonts}
\usepackage[noend]{algpseudocode}
\usepackage{graphicx}
\usepackage{textcomp}
\usepackage{float}
\usepackage{listings}
\usepackage{amsmath}
\usepackage{xspace}
\usepackage{multirow}
\usepackage{amsthm}

\usepackage{balance}
\usepackage{algorithm}
\usepackage{algpseudocode}
\usepackage{colortbl}

\usepackage[skins]{tcolorbox}
\usepackage{xcolor,pifont}
\usepackage{multicol}
\newcommand*\colourcheck[1]{%
	\expandafter\newcommand\csname #1check\endcsname{\textcolor{#1}{\ding{52}}}%
}
\colourcheck{blue}
\colourcheck{green}
\colourcheck{red}
\newtcolorbox{boxB}[2][]{%
  enhanced,colback=white,colframe=black,coltitle=black,
  sharp corners,
  toprule=1.0pt,
  rightrule=0.3pt,
  leftrule=0pt,
  bottomrule=0pt,
  fonttitle=\itshape\scshape\large,
  left=0pt,right=5pt,top=5pt,bottom=3pt,
  attach boxed title to top right={yshift=-0.3\baselineskip-0.4pt,xshift=-5mm},
  boxed title style={tile,size=minimal,left=0.2mm,right=0.5mm,
    colback=white,before upper=\strut},
  title=#2,#1
}

\newcommand{\tool}{\textsc{CodeFlow}\xspace}

\newboolean{showcomments}
\setboolean{showcomments}{true}
\ifthenelse{\boolean{showcomments}}
 { \newcommand{\mynote}[2]{
      \fbox{\bfseries\sffamily\scriptsize#1}
        {\small$\blacktriangleright$\textsf{\emph{#2}}$\blacktriangleleft$}}}
        { \newcommand{\mynote}[2]{}}

\usepackage{tikz}

\newtheorem{Definition}{Definition}

\newcolumntype{L}[1]{>{\raggedright\arraybackslash}p{#1}}

\newcommand{\code}[1]{{\footnotesize\texttt{#1}}}
\usepackage{amsthm}
\definecolor{dkgreen}{rgb}{0,0.6,0}
\definecolor{gray}{rgb}{0.5,0.5,0.5}
\definecolor{lightgray}{rgb}{211, 211, 211}
\definecolor{mauve}{rgb}{0.58,0,0.82}
\definecolor{c1}{HTML}{f4cccc}
\definecolor{c2}{HTML}{f5cdcd}
\definecolor{c3}{HTML}{fffcfc}
\definecolor{c4}{HTML}{ffffff}
\definecolor{c5}{HTML}{ffffff}
\definecolor{c6}{HTML}{fffdfd}
\definecolor{c7}{HTML}{f5cfcf}
\definecolor{c8}{HTML}{fffbfb}
\definecolor{c9}{HTML}{ffffff}
\definecolor{c10}{HTML}{fffdfd}
\definecolor{c11}{HTML}{fefafa}
\definecolor{c12}{HTML}{fef7f7}
\definecolor{c13}{HTML}{ffffff}
\definecolor{c14}{HTML}{fffefe}
\definecolor{c15}{HTML}{ffffff}
\definecolor{c16}{HTML}{fefafa}
\definecolor{c17}{HTML}{fdf3f3}
\definecolor{c18}{HTML}{fffefe}
\definecolor{c19}{HTML}{fdf5f5}
\definecolor{c20}{HTML}{ffffff}

\usepackage{tikz}
\usepackage{listings}
\usepackage{color}
\usepackage{float}
\usepackage{adjustbox}
\usepackage{amsmath,amssymb,amsfonts}
\usepackage{amstext}
\usepackage{xcolor}
\def\BibTeX{{\rm B\kern-.05em{\sc i\kern-.025em b}\kern-.08em
    T\kern-.1667em\lower.7ex\hbox{E}\kern-.125emX}}
\begin{document}

\title{{\tool}: Program Behavior Prediction with Dynamic Dependencies Learning}

\author{
    \IEEEauthorblockN{
        \begin{minipage}[t]{0.33\textwidth}
            \centering
            Cuong Chi Le\\
           \textit{FPT Software AI Center, Viet Nam}\\
            cuonglc4@fpt.com
        \end{minipage}
        \hfill
        \begin{minipage}[t]{0.33\textwidth}
            \centering
            Hoang Nhat Phan\\
            \textit{Nanyang Technological University, Singapore}\\
            c210055@e.ntu.edu.sg
        \end{minipage}
        \hfill
        \begin{minipage}[t]{0.33\textwidth}
            \centering
            Huy Nhat Phan\\
            \textit{FPT Software AI Center, Viet Nam}\\
            huypn16@fpt.com
        \end{minipage}
    }
    \\
    \IEEEauthorblockN{
        \begin{minipage}[t]{0.33\textwidth}
            \centering
            Tien N. Nguyen\\
            \textit{University of Texas at Dallas, USA}\\
            tien.n.nguyen@utdallas.edu
        \end{minipage}
        \begin{minipage}[t]{0.33\textwidth}
            \centering
            Nghi D. Q. Bui\\
            \textit{FPT Software AI Center, Viet Nam}\\
            nghibdq@fpt.com
        \end{minipage}
    }
}

\maketitle

\begin{abstract}
Predicting program behavior without execution is a critical task in software engineering. Existing models often fall short in capturing the dynamic dependencies among program elements. To address this, we~present {\tool}, a novel~machine learning-based approach that predicts code coverage and detects runtime errors by learning both static and dynamic dependencies within the code. By using control flow graphs (CFGs), {\tool} effectively represents all possible execution paths and the statistic relations between different statements, providing a more comprehensive understanding of program behaviors. 
{\tool} constructs CFGs to represent possible execution paths and learns vector representations (embeddings) for CFG nodes, capturing static control-flow dependencies. Additionally, it learns dynamic dependencies by leveraging execution traces, which reflect the impacts among statements during execution. This combination enables {\tool} to accurately predict code coverage and identify runtime errors. Our empirical evaluation demonstrates that {\tool} significantly improves code coverage prediction accuracy and effectively localizes runtime errors, outperforming state-of-the-art models.
\end{abstract}

\section{Introduction}
\label{sec:intro}

Large language models (LLMs) excel in understanding source code and descriptive texts~\cite{wang2023codet5, li2023starcoder, roziere2023code, nijkamp2023codegen2, luo2023wizardcoder,  mishra2024granite,  manh2023vault}. Their ability to recognize patterns, syntax, and semantics makes them effective at tasks such as code completion, bug detection, and generating human-readable explanations. However, state-of-the-art LLMs~\cite{tufano2023predicting,liu2023code} 
exhibit weaknesses in predicting dynamic program behavior, such as code coverage prediction and run-time error detection, etc, which typically require a program executable, but ideally, we want a model to predict them correctly without execution. This limitation arises from their reliance on static code representations, which fail to capture {\em dynamic program behavior} and state changes at runtime. Consequently, the models’ token-based predictions result in a superficial understanding of code, lacking context for variable states and control flow across multiple iterations. As a result, they struggle to accurately simulate loops, conditional branches, and the cumulative effects between statements. 
This shortcoming is further exacerbated by their inability to understand dynamic dependencies and interactions between various statements, making them ill-equipped to handle intricate control flows.

To address these limitations, several approaches have been proposed. The pre-trained model {\em TRACED}~\cite{ding2023traced} relies exclusively on the final execution of the last line within a loop to finalize the program states via variable value ranges, which leads to inadequate handling of condition and iteration statements. TRACED employs a variable coverage learning approach, labeling variable occurrences within an executed line. This may fail to capture branching behavior in scenarios where a branch lacks variable occurrences (e.g., having statements like \code{return}, \code{exit}, etc.) or in cases where a variable occurrence in a \code{true} branch occurs in one iteration but not in another. 
In contrast, CodeExecutor~\cite{liu2023code}, uses UniXcoder~\cite{guo-etal-2022-unixcoder} on pre-training data including the source code, input values, and the full execution trace with values at each execution step. It heavily relies on UniXcoder to transform the source code and its input into the entire sequence for the execution trace.

Toward dynamic program behavior prediction, we introduce {\tool}, a code coverage prediction model designed to predict code coverage given source code and its input. {\tool} leverages a control flow graph (CFG) that helps it better understand and predict the dynamic dependencies in code, including the execution of different branches and loops under varying conditions. With CFGs, we model loops as circular paths, allowing messages to pass through all possible paths and return to the loop node. This captures the aggregate effect of all iterations, ensuring the model understands cumulative changes in variables. To enhance coverage prediction, we focus on {\em learning dynamic dependencies via execution paths on CFG with respect to input values}. The CFG provides a detailed representation of the execution flows, capturing intricate paths through sequential, branching, and iterations.


To show  {\tool}'s usefulness in analyzing code coverage and dynamic behaviors of (in)complete code snippets,
we use it to build a tool to {\em statically detect runtime errors in both (in)complete snippets}. Platforms like Stack~Overflow (S/O) are invaluable resources for developers facing technical issues. However, S/O code snippets may contain hidden defects, runtime errors, and security vulnerabilities, posing potential risks to applications that integrate them~\cite{verdi-tse22,hong21dicos,fisher17stack,chaiyong2021toxic}. 
It is essential to directly analyze online code snippets to reason on their behaviors. Such vulnerabilities can crash processes or pose security risks if executed without prior analysis.
{\em The rationale for early detection is that the execution of such adapted code is unsafe due to the presence of pre-existing vulnerabilities in online code}. However, the key challenge is the incompleteness of online code snippets. This incompleteness may also arise from issues such as incompatible libraries or version mismatches (e.g., CUDA incompatibility), which prevent direct execution. Thus, predicting code coverage for incomplete code thus becomes a safer alternative. To do so, we use an LLM to act as a fuzzer, generating inputs to detect runtime errors in a given code snippet. Each input is used in the code snippet, which is then fed into {\tool} to predict the code coverage. If the code coverage stops unexpectedly and never reaches an exit point, {\tool} will locate the error.

We conducted an empirical evaluation on {\tool}. Our findings indicate that it significantly improves code coverage prediction, runtime error detection, and bug localization compared to existing models. Specifically, {\tool} achieves an accuracy of 75.24\% in matching code coverage exactly, outperforming GPT-4o at 68.13\%. For branch coverage, {\tool} reaches 87.88\%, significantly higher than GPT-4o's 78.75\%. In runtime error detection, {\tool} attains a high accuracy of 97.51\%, exceeding the performance of other models. Moreover, {\tool} maintains high accuracy even on incomplete code snippets, demonstrating its generalization capability. {\tool} also proves highly effective in supporting fuzz testing, particularly in scenarios involving incomplete code snippets where traditional execution is not feasible.


In brief, this paper makes the following contributions:

\begin{itemize}[leftmargin=*]
    \item \textbf{{\tool}: Dynamic Dependencies Learning for Code Coverage Prediction}: A novel code coverage prediction model leveraging CFGs to capture both static and dynamic code dependencies. {\tool} models loop as circular paths and learning dynamic dependencies among statements.
    \item \textbf{Effective Runtime Error Detection and Localization}: {\tool} analyzes code coverage continuity within CFGs to accurately detect and localize runtime errors.
    
    \item \textbf{Comprehensive Empirical Evaluation}: Experiments show that {\tool} outperforms existing models in code coverage prediction, runtime error detection, and localization.
\end{itemize}

\section{Motivation}
\label{sec:motiv}

\subsection{Example and Observations}

\definecolor{mygreen}{rgb}{0,0.6,0}
\definecolor{mygray}{rgb}{0.5,0.5,0.5}
\definecolor{mymauve}{rgb}{0.58,0,0.82}

\begin{figure}[t]
    \centering
    \begin{adjustbox}{width=0.48\textwidth}
    \begin{tikzpicture}
        \draw[black] (-4, 4.4) rectangle (9.2, -4.4);

        \node at (-2.6,4) {\textbf{True}};
        \node at (-0.8,4) {\textbf{CodeExecutor}};
        \node at (1,4) {\textbf{LLM}};
        \node at (4,4) {\textbf{Code Snippet}};
        
        \node[anchor=west] at (-4,3.5) {1};
        \node[anchor=west] at (-2.9,3.5) {$>$};
        \node[anchor=west] at (-1.1,3.5) {$>$};
        \node[anchor=west] at (0.7,3.5) {$>$};
        \node[anchor=west] at (3,3.5) {\lstinline|x = 10|};

        \node[anchor=west] at (-4,3) {2};
        \node[anchor=west] at (-2.9,3) {$>$};
        \node[anchor=west] at (-1.1,3) {$>$};
        \node[anchor=west] at (0.7,3) {$>$};
        \node[anchor=west] at (3,3) {\lstinline|while x > 4:|};

        \node[anchor=west] at (-4,2.5) {3};
        \node[anchor=west] at (-2.9,2.5) {$>$};
        \node[anchor=west] at (-1.1,2.5) {$>$};
        \node[anchor=west] at (0.7,2.5) {$>$};
        \node[anchor=west] at (3,2.5) {\lstinline|    if x % 2 == 0:|};

        \node[anchor=west] at (-4,2) {4};
        \node[anchor=west] at (-2.9,2) {$>$};
        \node[anchor=west] at (-1.1,2) {$>$};
        \node[anchor=west] at (0.7,2) {$>$};
        \node[anchor=west] at (3,2) {\lstinline|        print('x is even')|};

        \node[anchor=west] at (-4,1.5) {5};
        \node[anchor=west] at (-2.9,1.5) {$>$};
        \node[anchor=west] at (-1.1,1.5) {$>$};
        \node[anchor=west] at (0.7,1.5) {$>$};
        \node[anchor=west] at (3,1.5) {\lstinline|        x -= 2|};

        \node[anchor=west] at (-4,1) {6};
        \node[anchor=west] at (0.7,1) {$>$};
        \node[anchor=west] at (3,1) {\lstinline|    else:|};

        \node[anchor=west] at (-4,0.5) {7};
        \node[anchor=west] at (0.7,0.5) {$>$};
        \node[anchor=west] at (3,0.5) {\lstinline|        print("x is odd")|};

        \node[anchor=west] at (-4,0) {8};
        \node[anchor=west] at (0.7,0) {$>$};
        \node[anchor=west] at (3,0) {\lstinline|        x -= 1|};

        \node[anchor=west] at (-4,-0.5) {9};
        \node[anchor=west] at (-2.9,-0.5) {$>$};
        \node[anchor=west] at (0.7,-0.5) {$>$};
        \node[anchor=west] at (3,-0.5) {\lstinline|for i in range(100):|};

        \node[anchor=west] at (-4,-1) {10};
        \node[anchor=west] at (-2.9,-1) {$>$};
        \node[anchor=west] at (0.7,-1) {$>$};
        \node[anchor=west] at (3,-1) {\lstinline|    x += i|};

        \node[anchor=west] at (-4,-1.5) {11};
        \node[anchor=west] at (-2.9,-1.5) {$>$};
        \node[anchor=west] at (0.7,-1.5) {$>$};
        \node[anchor=west] at (3,-1.5) {\lstinline|if x % 3 == 0:|};

        \node[anchor=west] at (-4,-2) {12};
        \node[anchor=west] at (3,-2) {\lstinline|    print("x devide by 3 is 0")|};

        \node[anchor=west] at (-4,-2.5) {13};
        \node[anchor=west] at (-2.9,-2.5) {$>$};
        \node[anchor=west] at (3,-2.5) {\lstinline|elif x % 3 == 1:|};

        \node[anchor=west] at (-4,-3) {14};
        \node[anchor=west] at (-2.9,-3) {$>$};
        \node[anchor=west] at (3,-3) {\lstinline|    print("x devide by 3 is 1")|};

        \node[anchor=west] at (-4,-3.5) {15};
        \node[anchor=west] at (0.7,-3.5) {$>$};
        \node[anchor=west] at (3,-3.5) {\lstinline|else:|};

        \node[anchor=west] at (-4,-4) {16};
        \node[anchor=west] at (0.7,-4) {$>$};
        \node[anchor=west] at (3,-4) {\lstinline|    print("x devide by 3 is 2")|};
    \end{tikzpicture}
    \end{adjustbox}
    \caption{Code Coverage Prediction Comparison}
    \label{fig:code_snippet}
\end{figure}

Let us use an example to explain the problem and motivate our proposed solution. Fig.~\ref{fig:code_snippet} shows an example in Python with the input value $x$ = 10 and the code coverage, where "$>$" indicates the lines of code that are executed during actual execution. We employed two state-of-the-art approaches in CodeExecutor~\cite{liu2023code} and GPT-4~\cite{tufano2023predicting} to predict code coverage for our example and the results are shown in Fig.~\ref{fig:code_snippet}. 


\subsubsection{Observation 1. Conditional Statements}
Conditional statements (\code{if-elif-else}) present a challenge for existing approaches. For instance, after a \code{For} loop, the program checks the value of $x$. LLMs may skip necessary checks, resulting in incorrect predictions, such as jumping directly to the \code{else} statement and bypassing \code{elif}. This occurs due to their lack of state tracking across lines of code and inability to understand the dependencies between nested branches.

\subsubsection{Observation 2. Complex Loop Branching}

Loops, such as \code{while}, contain multiple branches determined by intermediate values of variables, leading to various outcomes such as skipping, entering, exiting, or continuing the loop. These values can change during the loop's iterations, making accurate prediction difficult with a top-to-bottom approach. For example, in the code snippet (Fig. \ref{fig:code_snippet}), the \code{while} loop on line 2 processes even and odd values of \code{x} differently. 
GPT-4 struggles to grasp the nuances of loop execution due to their reliance on token-based predictions without understanding the dynamic dependencies among statements via the state changes. 


CodeExecutor~\cite{liu2023code} correctly skips the \code{else} branch by capturing the intermediate values of \code{x} throughout the execution trace. However, it performs poorly with complex code requiring multiple iterations to update variable values. Error propagation frequently occurs, leading to incorrect coverage predictions.

\begin{figure}[!t]
    \centering
    \includegraphics[width=0.8\columnwidth]{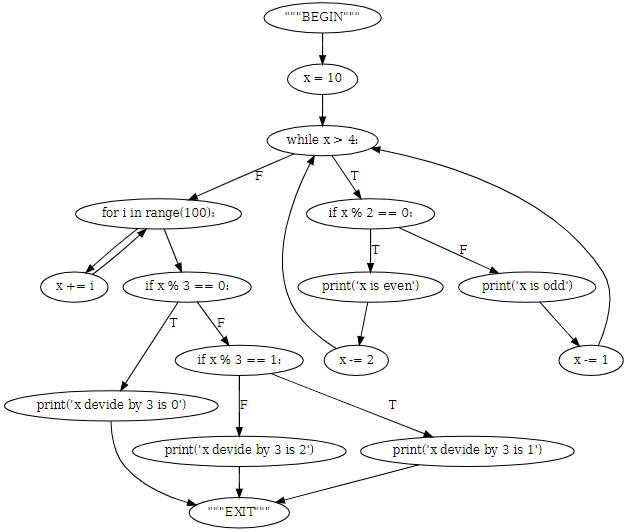}
    \caption{Control Flow Graph for code in Fig. \ref{fig:code_snippet}}
    \label{fig:cfg}
\end{figure}

\subsubsection{Observation 3: Information Loss in Repeated Loops}

For loops running several iterations, predicting the outcome based on line-by-line variable states often leads to incorrect results. For example, the loop on line 11 runs 100 iterations, updating \code{x} each time (lines 12-13). While the LLM correctly predicts that the loop will execute all iterations, it fails to understand the cumulative effect on \code{x}. This is because the LLM only processes static information, whereas the \code{for} loop requires information from the last line in the body to be fed back into the loop. 
After the loop, the value of \code{x} is used to decide which branch of the \code{if-elif-else} statement (lines 14-19) will be executed. The LLM incorrectly predicts the outcome because it does not account for the aggregated change in \code{x}. Consequently, it fails to accurately simulate the dynamic changes in variable states over multiple iterations. 


\subsubsection{Observation 4: Runtime Error Detection}
Detecting runtime errors requires understanding both the static and dynamic dependencies within the code. To determine whether a line contains a runtime error, it is crucial to know which lines are related and affect it (static dependencies). Additionally, understanding how variable changes impact the execution flow (dynamic dependencies) is essential. Existing models struggle with this task because they often fail to capture these intricate dependencies. They do not adequately analyze how changes in variable states influence subsequent lines of code, leading to missed detections. This lack of comprehensive dependency analysis makes it challenging for these models to pinpoint the exact line causing the runtime error and understand its context.



\subsection{Key Ideas}

From the above observations, we design our solution {\tool} with the following design strategies:

\subsubsection{Key Idea 1. [Learning Code Execution on Control Flow Graph]} Instead of reasoning the predicted execution on source code, we leverage a graph-based representation for such prediction and code coverage prediction: Control Flow Graph. 

\begin{Definition}[Control Flow Graph - CFG]
A control flow graph (CFG) is a graphical representation of the
control flow within a program. Nodes in the graph represent basic
blocks of code, such as individual or groups of statements
that are executed sequentially, while edges represent the flow of
control between these blocks, typically based on conditions such as
loops, conditional statements (e.g., \code{if-else}), or function
calls.
\end{Definition}

Fig. \ref{fig:cfg} displays the corresponding CFG of the code in Fig.~\ref{fig:code_snippet}. The CFG illustrates the sequence of execution of {\em statements} or {\em code blocks} within a program and {\em the conditions} that decide the control flow between different blocks, which are divided according to the program semantics. Learning 
execution on CFG provides several benefits.
First, CFGs (Fig.~\ref{fig:cfg}) explicitly represent the sequential nature of condition checks, ensuring all paths are considered and the model can accurately predict the execution flow based on all possible conditions. 
%
Second, training a model on code coverage using a CFG offers significant advantages over training on source code alone. This allows the model to better understand and predict the dynamic behavior of code, including how different branches and loops are executed based on varying conditions. In contrast, source code only provides static information without context on how the execution evolves.
Third, using CFGs to model loops as circular paths, allowing messages to pass through all possible paths and return to the loop node. This method captures the aggregate effect of all iterations, ensuring the model~comprehensively captures the cumulative changes in variables.

\subsubsection{Key Idea 2. [Dynamic Dependencies Learning via Execution Paths on CFG]}
A CFG is like a map that provides a blueprint for all possible paths, while an execution path is like a specific travel route on that map, tailored to a particular input of the program. To better predict code coverage, we aim to {\em learn dynamic dependencies among statements on CFG via a large number of execution paths} with respect to different inputs. That allows a model to better learn the representations of the execution flows, capturing the dynamic dependencies through sequential, branching, and iteration statements.

%


\subsubsection{Key Idea 3. [Detecting Runtime Error via CFG]} Once we have the code coverage, we combine it with the static dependency information between each line of the CFG to detect runtime errors. By checking the continuity in code coverage on the CFG, we can effectively identify a runtime error where the execution path unexpectedly terminates. Specifically, if the predicted path does not reach the \textit{EXIT} node, we trace back to the furthest node reached without an outgoing edge on the CFG. This node is likely to be the crash point and hence contains the error. By ensuring that the model predicts a continuous path in the CFG, we can accurately detect runtime errors and precisely localize the line containing the bug. 





\section{Approach Overview}
\label{sec:overviews}

\begin{figure*}[!t]
    \centering
    \includegraphics[width=\textwidth]{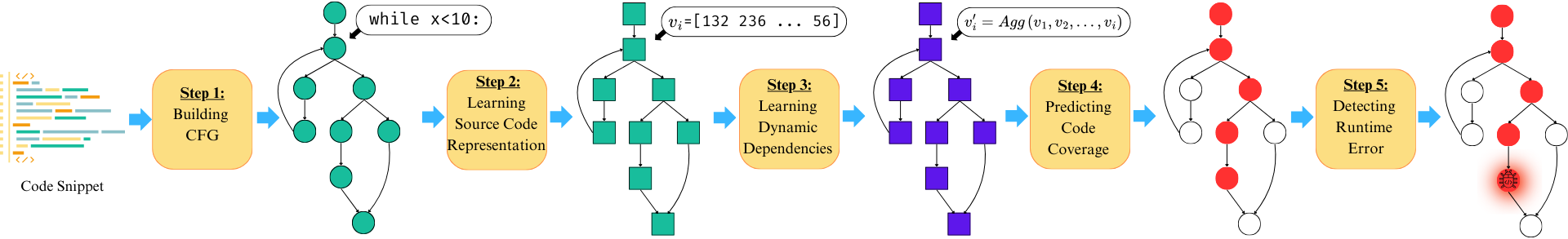}
    \caption{{\tool}: Predictive Code Coverage and Runtime Error Detection with Dynamic Dependencies Learning on CFG}
    \label{fig:overview}
\end{figure*}


Putting together our above ideas, we develop {\tool}, a code coverage prediction model that is given a source code and its input and predicts the corresponding code coverage.






Fig.~\ref{fig:overview} illustrates {\tool}'s overall architecture. The input is the source code that needs to be predicted.

\subsubsection*{Step 1. CFG Building (Section~\ref{sec:cfg})} First, the given source code is parsed to build the Abstract Syntax Tree (AST) and the CFG. The input of the source code is encoded as the assignments of the input variables with their values. Additionally, we apply processing steps to generalize and standardize the CFG including 
normalization of node labels, removal of redundant nodes, and simplification of complex structures.


\subsubsection*{Step 2. Source Code Representation Learning (Section~\ref{sec:node_embedding})} The goal of this module is to learn the vector representations (embeddings) for the nodes in the CFG that takes into account the {\em static} control-flow dependencies between the statements represented by the connected nodes. We used a Gated Recurrent Unit (GRU) Networks to transform the code into the embeddings that preserve the contextual and semantic
information. The output of this step is the CFG structure with each node represented by its corresponding embedding.




\subsubsection*{Step 3. Dynamic Dependencies Learning  (Section~\ref{sec:dynamic})} The goal of this step is to learn the dynamic dependencies that pertain to the execution of two connected nodes/statements. Let us call them {\em execution-based dynamic dependencies} or {\em dynamic dependencies} for short, which indicates whether a statement  (represented by a node) would be executed if a node/statement connected with that node is executed. To teach our model on such dynamic dependencies, we use the actual execution traces for the source code and inputs in the training data. We leverage a specialized message-passing scheme with a binary soft decision branching technique to effectively learn the interactions and dependencies that influence code coverage. The output of this step is the CFG structure with its nodes represented by the new vector representations that {\em capture dynamic dependencies during execution}.


\subsubsection*{Step 4. Code Coverage Prediction via Classification (Section~\ref{sssec:classification})} The goal of this step is to predict the code coverage for the statements or branches in the given code. Specifically, we use the learned embeddings from Dynamic Dependencies Learning to classify whether a specific node or branch will be covered during actual code execution. 

\subsubsection*{Step 5. Runtime Error Detection and Localization (Section~\ref{sssec:detection})} Finally, we use the code coverage predictions from Step 4 to detect whether the code contains runtime errors or not. Moreover, by analyzing the predicted code coverage along with the CFG, we identify nodes where the execution unexpectedly terminates, indicating a potential runtime error.


\section{Control Flow Graph Building}
\label{sec:cfg}

In the initial step, we create a CFG from a given code snippet to capture the static dependencies between different code blocks. Fig.~\ref{fig:cfg} illustrates the CFG of the code shown in Fig.~\ref{fig:code_snippet}. However, the original CFG often contains redundant information and lacks clarity in certain nodes, such as those representing loop conditions. To ensure consistency in loop representation and make the CFG easier to process and learn on, we convert \texttt{for} loops into \texttt{while} loops, treating them as condition nodes. This transformation helps maintain only two types of nodes in the graph: \textbf{operation nodes}, which have only one outgoing edge, representing a {\em sequential order} of statements, and \textbf{condition nodes}, which have more outgoing edges, representing {\em branching} based on conditions, simplifying the embedding process. Additionally, we enhance CFG's clarity by removing redundant information from each node, such as condition symbols (\code{if}, \code{elif}, \code{else}, and \code{while}), and adding markers \textbf{T} to distinguish \textit{True} and \textit{False} branches of condition statements. These modifications ensure each node has a uniform structure, making it easier for our model to learn and capture comprehensive information from the code.

In the end of the process, given initial program $P$ will be broken down into graph $G=(V, E)$ with set of node $V=\{n_1, n_2, \cdots, n_L\}$ with $L \geq 1$ as the number total of nodes. And the edge set $E$ consists of two edge types: forward edge and backward edge. In addition, each node $n_i$ also consists of a code statement to represent the semantic information.

\section{Source Code Representation Learning}
\begin{figure}[!h]
    \centering
    \includegraphics[width=1.2in]{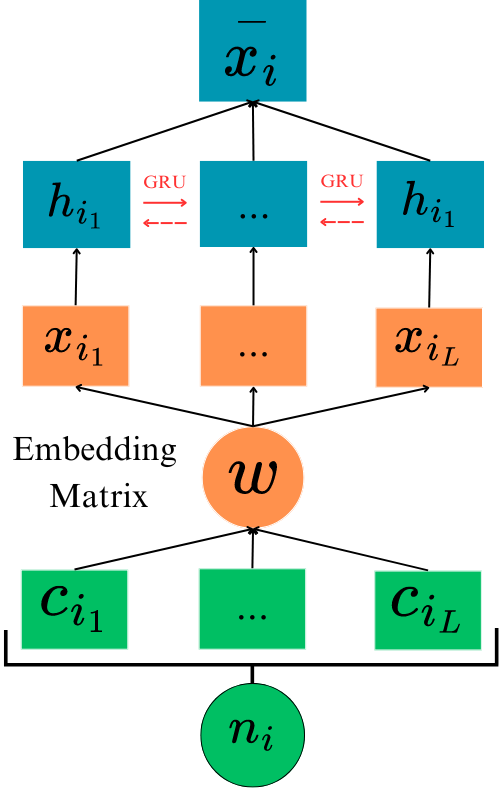}
    \vspace{-6pt}
    \caption{Source Code Representation Learning}
    \label{fig:arch}
\end{figure}
\label{sec:node_embedding}
The fundamental part of modeling the dynamic execution of a program is to statistically analyze the interactions of statements. Capturing the semantic information of those statements is the first stage of almost all machine learning (ML) approaches toward treating code as a sequence of tokens. Following this direction, we treat the node's statement $n_i$ as sequence of lexically tokens $c_{i_1}, c_{i_2}, \cdots, c_{i_L}$. Each token is then embedded into a vector $x_{i_t} = W_e c_{i_t}$ using a randomly initialized embedding matrix $W_e$. $W_e$ is a learnable parameter, a part of our training end-to-end system. Since in our scope of experiments, each program $P$ is frequently broken down to fine-grained short repeatedly constituents $n_i$. Thus, normally a node $n_i$ often consists of a short sequence of tokens. For this reason, we employ a much simpler model, GRU \cite{chung2014empiricalevaluationgatedrecurrent}, which is much simpler than state-of-the-art or frequently used recurrent models like Transformer \cite{vaswani2023attentionneed}, or LSTM \cite{10.1162/neco.1997.9.8.1735}. GRU still employs the gate mechanism - a mechanism to model long dependency tokens interaction, similar to LSTM. But by dropping unnecessary forget gates as in LSTM, we reduce the number of parameters, improving training efficiency and less prone to overfitting in the case of our experiments.

Code tokens relation is not increasing order, left to right manner as in natural language but rather in both directions of appearance. We calculate the node embedding $\overline{x_i}$ of $n_i$ via each token embeddings $x_{i_1}, x_{i_2}, \cdots, x_{i_L}$ as follows:
\begin{equation}
    h_{i_t} = GRU(x_{i_t}, h_{i_{t-1}})\quad \text{for}\, \, t = \{2, \cdots L\} \\
\end{equation}
\begin{equation}
    \overline{x_i} = Average(h_{i_1}, \cdots, h_{i_L})
\end{equation}
Element-wise average pooling operation $Average$ aggregates the state of a token to form final embedding $\overline{x_i}$ of the node~$n_i$.

\section{Dynamic Dependencies Learning}
\label{sec:dynamic}

\begin{figure}[!h]
    \centering
    \includegraphics[width=3.3in]{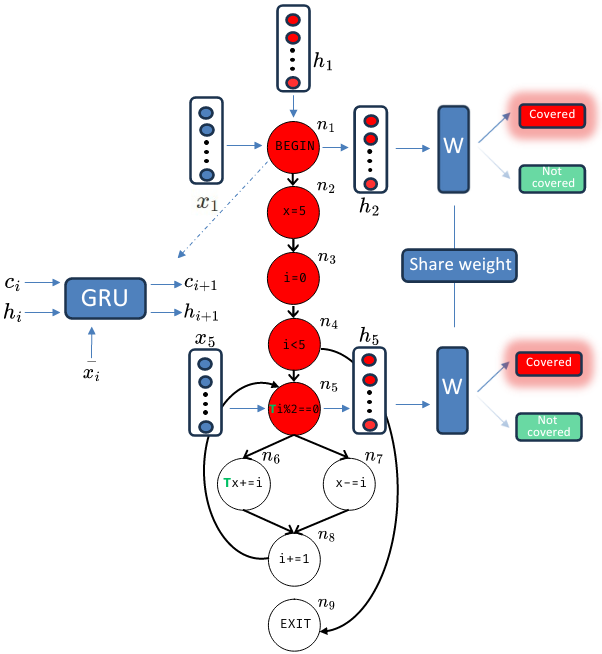}
    \vspace{-6pt}
    \caption{Dynamic Dependencies Learning}
    \label{fig:arch}
\end{figure}
A successful execution in CFG is represented by a path $P$ with order starting from the ``BEGIN'' node denoted as $n_0$, and sequentially adds node $n_i$ until ``EXIT'' node. To check the coverability of one node $n_j$, we need to check if $n_j$ appears in $P$ or not. We can treat this problem as a binary classification task with classification label $t_i$=$1$ meaning node $n_i$ is covered and $t_i=0$ as node $n_i$ is not appearing in the execution path: $t_i \sim P(t|\{n_{j}\}_{j<i})$.
However, not all the previous nodes in BFS are needed for the conditioning. But rather depends on the CFG itself. We represent the state of execution of node $n_i$ by $h_i \in \mathbb{R}^{n}$ with $h_0 = \bold{0}$ at the beginning of the execution. Considering an execution path $P=\{n_{i_1}, n_{i_2},\cdots, n_{i_k}\}$ of length $k$ with $i_1 < i_2 < \cdots < i_k$. {\em The probability of appearance of a current node $i_j$ is conditioned on previous nodes}: $t_{i_j} \sim P(t|n_{i_1}, n_{i_2}, \cdots, n_{i_{j-1}})$.
One might consider using recent recurrent networks like traditional $RNN$ or $LSTM$ to model this conditioned distribution. With node embeddings $\overline{x_1}, \overline{x_2}, \cdots$ are computed in previous steps, we update the state $h_i$ by following equations:

\begin{eqnarray} \label{eq:aggregate}
  h^{*}_i     &= f(h_0, h_1, \cdots, h_{i-1}) \\
  c^{*}_i     &= f(c_0, c_1, \cdots, c_{i-1})
\end{eqnarray}
\vspace{-23px}
\begin{eqnarray}
  \text{in}_i &= \sigma ( W_{\text{in}}\overline{x_i} + U_{\text{in}}h^{*}_i + b_{\text{in}} ) \\
  \text{fg}_i &= \sigma ( W_{\text{fg}}\overline{x_i} + U_{\text{fg}}h^{*}_i + b_{\text{fg}} ) \\
  \text{op}_i &= \sigma ( W_{\text{op}}\overline{x_i} + U_{\text{op}}h^{*}_i + b_{\text{op}} ) \\
  \text{u}_i  &= \text{tanh} ( W_{\text{u}}\overline{x_i} + U_{\text{u}}h^{*}_i + b_{\text{u}} ) 
\end{eqnarray}
For computing the next state and memory cell $h_i$ and $c_i$:
\begin{eqnarray}
  c_i         &= \text{in}_i \odot u_i + \text{fg}_{i} \odot c^{*}_{i} \\
  h_i         &= \text{op}_i \odot \text{tanh} (c_i)
\end{eqnarray}
With function $f$ in \eqref{eq:aggregate} is aggregation function from previous states to summarize history information. For usual language modeling task, the function $f$ normally takes form of $f(h_0, h_1, \cdots, h_{i-1}) = h_{i-1}$.
But considering the following characteristics from our problem, we propose an adaptation to the original LSTM which was originally used for language modeling $P(t_i|t_0, t_1, \cdots , t_{i-1})$:

1. \textbf{CFG edges aggeregation:} in our CFG, a node $n_i$ has adjacency matrix considered only forward edges, denoted as $A_{\text{forward}}$. The aggregation function $f$ to be an average of adjacent nodes' states, $f(h_i) = A_{\text{forward}} H$ with $H = [h_0, \cdots, h_{i-1}, \bold{0}, \cdots, \bold{0}]$

2. \textbf{Forward and backward passing:} in our CFG, {\em a loop is broken down into condition node, body, and step node (e.g: $\text{counter}+=1$)}. The forward edges are in an increasing order, $n_i \to n_j$ but what is special is an additional backward edge from the loop step node to the condition~one. To~propagate the information from the step node $n_j$ to condition~node $n_i$ with $i$$<$$j$ but not to mix up the recurrent relations of execution. We update the state $h_i$ by $h_j$ for this special backward edge only by information of node embedding $\overline{x_j}$. By updating only the information but not the state at node $n_j$ which is not yet computed by forward order, we can combine both forward and backward propagation by only updating $h_i = LSTM(h_j)$ with $i$$<$$j$ instead of bidirectional like $BiLSTM$.

3. \textbf{Binary soft decision branching:} we have processed a condition (including loop condition) node $n_i$ will only connect to two nodes $n_j$ and $n_k$ with $i < j, k$ by two forward edges. And each of these edges is the only incoming edge to $n_j$ and $n_k$. The original computation of node $n_j$'s hidden state is $h_j^{*} = A_{forward} H = [h_0, \cdots, h_{i-1}, \bold{0}, \cdots, \bold{0}] = h_i$, similarly, $h_j^{*} = h_i$. With that, we allow information to pass to both possible branches while they are complemented in real code execution. Thus in order to model the branching behavior in the condition node, we will charge $A$ dynamically based on the current hidden state $h_i$. $A[i,j] = 0$ if $ Average(h_i) \geq 0$ and $A[i,k] = 0$ if $Average(h_i) < 0$ with assumption $j < k$. This will force the weights to adaptively produce reasonable $h_i$ to make a correct branch decision. A similar approach is taken by \cite{bieber2020learning}, but the major difference is that they add additional parameter complexity to learn this soft dynamic branching while we focus more on the efficiency by setting branching conditions depending only on the current hidden state.
\section{Coverage Prediction and Error Localization}
\label{sec:classification_detection}

\subsection{Coverage Prediction}
\label{sssec:classification}
In this step, we use the hidden states $h_i$ learned from Step~3 (Section~\ref{sec:dynamic}) to predict the code coverage. Each hidden state $h_i$ is passed through a linear layer followed by a sigmoid activation function to compute the coverage score. The score is then compared against a threshold $\alpha$ to classify whether a node is covered. Specifically, the process is defined as follows:

With $h_i$ $\in$ $\mathbb{R}^n$ be the final hidden state for node $n_i$~after dynamic dependencies learning, we compute the coverage~score:
\begin{equation}
s_i = \sigma(W_{c} h_i + b_{c})
\end{equation}
where $W_{c} \in \mathbb{R}^{1 \times n}$ and $b_{c} \in \mathbb{R}$ are the weights and bias of the linear layer, and $\sigma$ denotes the sigmoid activation function.

The coverage classification for node $n_i$ is determined by comparing $s_i$ to the threshold $\alpha$:
\begin{equation}
\hat{t}_i =
\begin{cases}
1 & \text{if } s_i \geq \alpha \\
0 & \text{otherwise}
\end{cases}
\end{equation}
where $\hat{t}_i$ is the predicted coverage label for node $n_i$. A value of $\hat{t}_i = 1$ indicates that node $n_i$ is predicted covered, while $\hat{t}_i = 0$ indicates that it is predicted to be covered, while $\hat{t}_i = 0$ indicates that it is predicted to be not covered.

To train our model, we use the Binary Cross-Entropy (BCE) loss function, which is suitable for binary classification tasks. The BCE loss for a single node $n_i$ is given by:
\begin{equation}
\mathcal{L}_i = -\left[t_i \log(s_i) + (1 - t_i) \log(1 - s_i)\right]
\end{equation}
where $t_i$ is the true label (1 if the node is covered, 0 otherwise) and $s_i$ is the predicted coverage score.

The total loss $\mathcal{L}$ over all nodes is the average of all losses:
\begin{equation}
\mathcal{L} = \frac{1}{N} \sum_{i=1}^{N} \mathcal{L}_i
\end{equation}
where $N$ is the total number of nodes in the training set.
 
\subsection{Runtime Error Detection and Localization}
\label{sssec:detection}
The underlying idea for runtime error detection is that code without runtime errors will terminate normally, covering both the \textit{BEGIN} and \textit{EXIT} nodes in the CFG. In contrast, buggy code will crash during execution, resulting in the \textit{EXIT} node not being reached. Therefore, we focus on the coverage of the \textit{EXIT} node to identify the presence of runtime errors.

One critical issue with existing models in predicting code coverage is the lack of continuity in the CFG. Discontinuity leads to gaps in the predicted execution path, making it difficult to accurately localize errors. {\tool}, addresses this issue by consistently predicting a concrete, continuous path from the \textit{BEGIN} to the \textit{EXIT} node. This continuity ensures that the predicted execution flow closely follows the actual control flow of the program.
Finally, to detect and localize runtime errors, we analyze the predicted code coverage as follows:

   \begin{itemize}
       \item \textbf{Runtime Error Check:} If {\em {\tool} predicts \textit{EXIT} node as a covered node, the code is likely free of runtime errors. If it does not, we infer that the code has crashed}.
       \item \textbf{Error Localization:} In the buggy code, {\em the furthest node reached without an outgoing edge is identified as the crash point, indicating the location of the runtime error}.
   \end{itemize}
   

By leveraging the continuity and comprehensive path prediction capabilities of {\tool}, we improve the reliability of error detection and localization.

\section{Empirical Evaluation}
\label{sec:empirical}

For evaluation, we seek to answer the following questions:

\textbf{RQ1. [Coverage Prediction Accuracy]}: How well does {\tool} predict code coverage for (in)complete code?


\textbf{RQ2. [Runtime Error Detection Accuracy]}: How well does {\tool} detect runtime errors in (in)complete code?

\textbf{RQ3. [Runtime Error Localization Accuracy]}: How accurately does {\tool} locate the lines with a runtime error?

\textbf{RQ4. [Usefulness in Fuzz Testing]}: How useful does {\tool} support fuzz testing in detecting runtime errors for (in)complete code snippets?

\begin{table*}[ht]
\centering
\caption{Code Coverage Prediction Comparison (RQ1).}
\label{tab:RQ1results}
\resizebox{0.8\textwidth}{!}{%
\begin{tabular}{l c c c c c c c c c c}
\toprule
 \multirow{2}{*}{\textbf{Model}} & \multicolumn{5}{c}{\textbf{Complete Code}} & \multicolumn{5}{c}{\textbf{Incomplete Code}} \\
\cmidrule(lr){2-6} \cmidrule(lr){7-11}
 & \textbf{EM (\%)} & \textbf{BC (\%)} & \textbf{P} & \textbf{R} & \textbf{F1} & \textbf{EM (\%)} & \textbf{BC (\%)} & \textbf{P} & \textbf{R} & \textbf{F1} \\
\midrule
CodeExecutor & 18.83 & 31.34 & 0.94 & 0.47 & 0.70 & 10.45 & 25.50 & 0.90 & 0.42 & 0.66 \\
CFGNN & 45.53 & 76.56 & 0.92 & 0.91 & 0.92 & 44.32 & 77.10 & 0.90 & 0.89 & 0.90 \\
\midrule
\multicolumn{11}{l}{\textbf{LLMs}} \\
\quad Gemini & 56.17 & 74.96 & 0.87 & 0.97 & 0.92 & 59.10 & 73.85 & 0.88 & 0.95 & 0.91 \\
\quad Claude & 64.94 & 77.30 & 0.96 & 0.94 & 0.95 & 66.50 & 79.00 & 0.95 & 0.93 & 0.94 \\
\quad GPT-4 & 68.13 & 78.75 & 0.96 & 0.96 & 0.96 & 67.75 & 80.20 & 0.96 & 0.95 & 0.96 \\
\midrule
\rowcolor{green!15} \textbf{\tool} & \textbf{75.24} & \textbf{87.88} & \textbf{0.97} & \textbf{0.97} & \textbf{0.97} & \textbf{76.50} & \textbf{86.95} & \textbf{0.96} & \textbf{0.98} & \textbf{0.97} \\
\bottomrule
\end{tabular}%
}
\end{table*}

\section{Code Coverage Prediction Accuracy (RQ1)}
\label{sec:RQ1}

\subsection{Data Collection, Baselines, Procedure, and Metrics}

\subsubsection{Datasets} For training, we utilize a comprehensive dataset specifically curated for code coverage prediction. Our primary dataset, CodeNetMut, is derived from Liu {\em et al.}~\cite{liu2023code}. This dataset was created by crawling and generating mutations based on submissions to competitive programming problems from the CodeNet dataset~\cite{puri2021codenetlargescaleaicode}. 
CodeNetMut contains nearly 20,000 Python files. After excluding those that failed execution by python-trace or CFG construction by python-graphs, we were left with 8,216 Python code snippets. 

However, CodeNetMut lacks a sufficient number of Python files with extensive conditional statements, which are crucial for training the model on conditional branching. To address this, we supplemented CodeNetMut with an additional dataset generated using Gemini-API. This synthetic dataset comprises approximately 11,668 Python code snippets, each featuring diverse and complex statements. The code snippets vary in size, with the largest containing 146 lines of code and a mean length of 13. 
Over 4,500 Python code snippets (23\%) have a Cyclomatic Complexity above 10, being classified as complex and challenging to test \cite{mccabe2010metrics}. For each snippet, we generate the CFG, tracking nodes, forward edges (normal control flow), and backward edges (loop control flow). The dataset was split 80:20 into training and testing sets. 


To build the ground truth in training, we use the \code{trace} library from Python to record 
the code coverage. In addition, we created the \textbf{Incomplete Code dataset} by removing all import statements and external file references from each snippet.



\subsubsection{Baselines and Procedure}
We compare {\tool} with the following approaches: 

1. \textbf{CodeExecutor}~\cite{liu2023code}: primarily predicts execution traces. It leverages the transformer-based UniXcoder model, which is trained via the data including source code, input values, and the full execution traces with values at each execution step. 




2. \textbf{CFGNN}~\cite{10.1145/3597926.3598142}: originally designed for detecting condition-related bugs via CFGs. We modified CFGNN by retaining its main architecture but altering the final linear layer to output a list of scores for each node, instead of a single node, allowing it to predict coverage across multiple nodes.

3. \textbf{OpenAI GPT-4o} (gpt-4o), \textbf{Anthropic Claude} (claude-3.5-sonnet), and \textbf{Google Gemini} (gemini-1.5-flash): We used several LLMs as baselines. We follow Tufano {\em et al.}~\cite{tufano2023predictingcodecoverageexecution}~to design the prompt to GPT-4 to get the code coverage.


\subsubsection{Evaluation Metrics}
We use the following key metrics. 


+ \textbf{Exact Matching (EM)}: This metric counts the number of times when the entire predicted sequence of statements exactly matches the target sequence of {\em true coverage}, representing the model's capability to predict the executed statements.

+ \textbf{Branch Coverage Matching (BC)}: This metric counts the number of times a model correctly predicts the {\em branch coverage} (at a condition node in CFG), assessing a model's prediction on conditions and loops.


+ \textbf{Precision (P), Recall (R), F1-Score (F1)}: These metrics are determined by consolidating all nodes in the test set into a unified dataset and calculating the metrics for that. Precision measures the proportion of nodes predicted as covered that were actually covered during execution, while recall reflects the proportion of executed nodes correctly predicted as covered. F1-Score is the harmonic mean of precision and recall.


\subsection{Empirical Results}


As seen in Table~\ref{tab:RQ1results}, {\tool} outperforms existing models across all metrics for both complete and incomplete code. 

\subsubsection{Exact-Matching} For complete code, it achieves an {\em exact matching accuracy} of \textbf{75.24\%}, which is higher than the best-performing LLM, GPT-4o, at 68.13\%. Notably, {\tool} accomplishes this with far fewer parameters (1.3 million in total) compared to LLMs like GPT-4o (over 1 trillion) and Claude (175 billion), which suggests a more efficient architecture for this task and underscores its practicality for the scenarios where computational resources may be limited. 

\subsubsection{Branch-coverage Matching} Our model achieves 87.9\% correctness, outperforming the best baseline, GPT-4o, by nearly 10\%. This highlights the efficacy of using CFGs to capture the complex behaviors of loops and conditional branches. 
With CFG modeling the intricate decision points within code, {\tool} understands the cumulative effect of variable changes over multiple iterations, leading to more accurate branch predictions. In contrast, models like CodeExecutor fail in this aspect because they do not adequately handle dynamic execution changes, leading to predictions that do not align well with actual execution paths involving loops and conditions.


{\tool} achieves 97\% in all three metrics, slightly surpassing the best LLMs (Claude and GPT-4o), which scored 96\%. Its high precision shows effectiveness in identifying executed lines while minimizing false positives, and its recall demonstrates strong ability to capture all executed lines without missing any. Notably, half of the test dataset includes solutions from the CodeNet Project, published online since 2022, potentially benefiting LLMs if trained on this data. Despite this, {\tool} still outperforms them.

\subsubsection{Incomplete Code}
To our knowledge, no runtime error dataset exists for incomplete Python code snippets paired with their complete versions (to build ground truth). To evaluate {\tool}'s capability with incomplete code—specifically without built-in and external library imports—we trained and tested it on a dataset where all import statements and method/class declarations were removed. This preserved the code's semantic integrity while retaining function/API calls. Despite the absence of imports, {\tool} outperformed other models, as shown in Table~\ref{tab:RQ1results}, due to its ability to learn library semantics during training and predict code behavior based on nodes using library functions. Similar performance was observed in LLMs (GPT-4o, Claude, and Gemini) due to their extensive pre-training on library semantics. CodeExecutor, which tracks intermediate execution values, and CFGNN, which models control flow, also performed well without import information.

\begin{figure*}[!t]
	\centering
	\small
	\includegraphics[width=\textwidth]{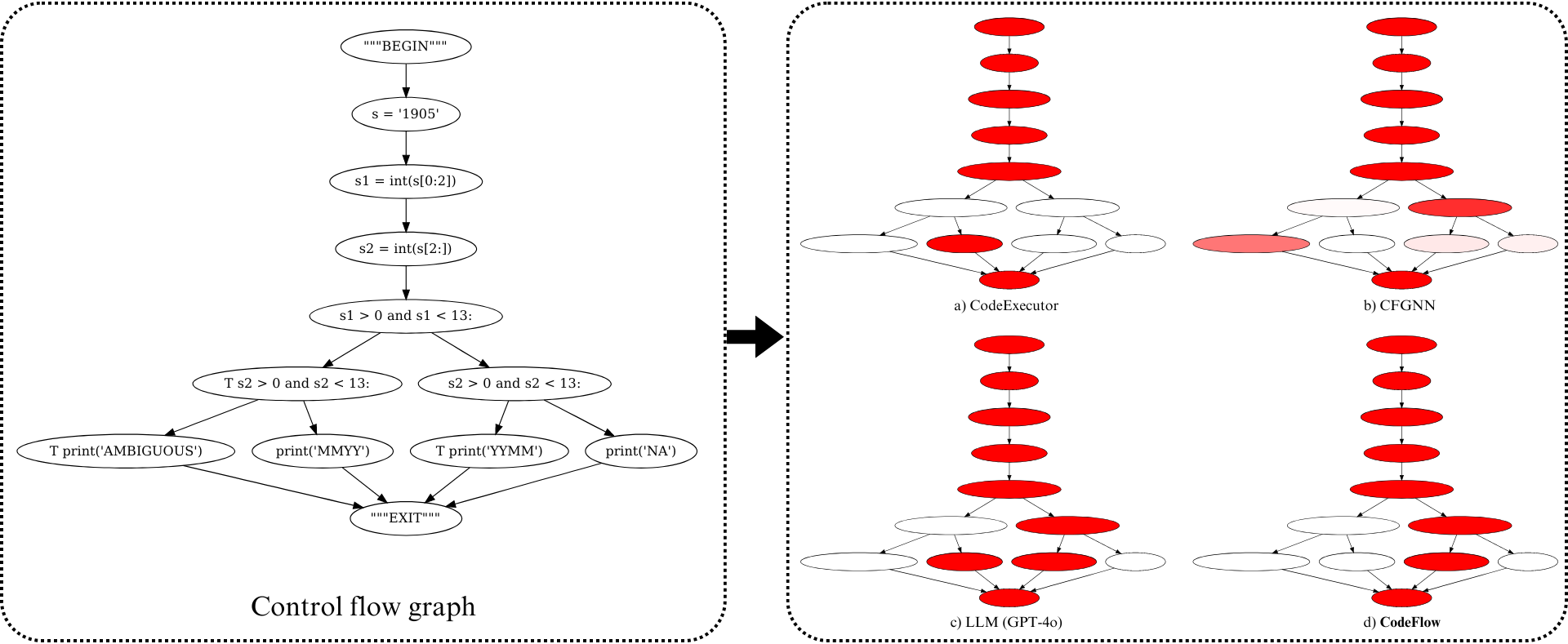}
	\caption{Code coverage prediction visualization for different models with red nodes indicate coverages. For CFGNN and \tool, the nodes' shades correspond to values from 0 to 1, representing the predicted coverage scores.}
	\label{fig:visualization}
\end{figure*}

\subsubsection{Continuity in Predicting Code Coverage} Continuity means a model should predict nodes or lines of code that are sequential or logically connected, rather than skipping intermediate nodes. This issue is common in existing methods. CodeExecutor, which relies on exact execution traces during training, often misses dependencies between lines. As shown in Fig.~\ref{fig:visualization}a, CodeExecutor skips a node and jumps directly to the next, creating a discontinuous CFG path. 

Similarly, CFGNN transmits information equally through all possible paths in the CFG, rather than focusing on the correct execution paths. This approach often leads to a misunderstanding of the continuity in code coverage tasks. Fig.~\ref{fig:visualization}b clearly shows the heatmap of predicted scores for each node, highlighting CFGNN's shortcomings in coverage prediction. 

LLMs like GPT-4o also struggle with continuity, often failing to capture dynamic relations between lines and skipping critical steps, such as the \code{elif} in an \code{if-elif-else} structure (Fig.~\ref{fig:visualization}c). This happens because LLMs rely on next-token probabilities, predicting the most likely token based on prior context. As a result, they may misinterpret code structure, especially dynamic behaviors and state changes across iterations. Consequently, despite high precision, recall, and F1 scores, their exact-matching accuracy is lower.

{\tool}, while similar to CFGNN in using CFGs, is designed to address this issue. By emphasizing the correct path through CFGs and ensuring that information is passed predominantly along the actual execution paths, {\tool} maintains continuity in its predictions (Fig.~\ref{fig:visualization}d). The model predicts sequences of executed lines that are connected, following a coherent path from the beginning to the end.

\section{Runtime Error Detection Accuracy (RQ2)}
\label{sec:RQ2}

In this study, we assess the ability of {\tool} to predict whether a given code snippet contains a runtime error.

\subsection{Data Collection and Evaluation Metrics}
We used a dataset in addition to that in~RQ1. Specifically, we used the FixEval dataset, which comprises 2,066 unique problems with 277,262 submissions of Python code snippets. From this dataset, test cases were obtained for 800 problems from the CodeNet dataset~\cite{puri2021codenetlargescaleaicode}. Each of these snippets, when executed with its respective input, leads to a runtime error.
After filtering, we obtained 6,437 submissions across the 800 problems. This combined dataset, referred to as the {\em Complete Runtime-Error Dataset}, includes {\em both code snippets that terminate normally and those that encounter runtime errors}. 



In addition to the metrics in RQ1,
we use \textbf{Runtime Error Detection Accuracy (EDA)} to measure the accuracy of a model correctly predicting if a snippet has an~error.

\subsection{Empirical Results}
\begin{table}[t]
\centering
\caption{Runtime Error Detection Comparison (RQ2).}
\label{tab:RQ2results}
\begin{tabular}{l c c c c}
\toprule
 \textbf{Model} & \textbf{EDA (\%)} & \textbf{P} & \textbf{R} & \textbf{F1} \\
\midrule
CFGNN & 76.71 & 0.51 & 0.89 & 0.65 \\
\midrule
\multicolumn{5}{l}{\textbf{LLMs}} \\
\quad Claude & 77.98 & \textbf{0.98} & 0.89 & 0.93 \\
\quad GPT-4o & 69.24 & 0.71 & \textbf{0.99} & 0.83 \\
\midrule
\rowcolor{green!15} \textbf{\tool} & \textbf{97.51} & 0.96 & 0.94 & \textbf{0.95} \\
\bottomrule
\end{tabular}
\end{table}
As seen in Table~\ref{tab:RQ2results}, Claude achieved the highest precision of 0.98, indicating its strong ability to correctly identify runtime errors when they are present. High precision means Claude makes very few false positive predictions, thus showing its accuracy in pinpointing real runtime errors. However, Claude's recall score of 0.89, while still respectable, is lower than that of GPT-4o. This suggests that Claude may miss some runtime errors, indicating that it is more conservative in error detection.

In contrast, GPT-4o achieved the highest recall of 0.99, showing its effectiveness in identifying nearly all runtime errors. GPT-4o's precision score of 0.71 indicates a higher rate of false positives compared to Claude, meaning it sometimes incorrectly flags non-buggy code snippets as erroneous. This suggests that GPT-4o could have more false alarms.

CFGNN shows a more moderate performance with a runtime error detection accuracy of 76.71\%, a precision of 0.51, a recall of 0.89, and an F1-score of 0.65. This implies that while CFGNN can detect errors, it struggles to accurately discriminate between erroneous and non-erroneous code snippets, leading to many false alarms. This performance is likely due to CFGNN's approach of transmitting information equally through all possible paths in the CFG, which might result in overestimating the likelihood of errors.

As seen, \textbf{{\tool}} exhibits a balanced performance with high scores across all metrics: a precision of \textbf{0.96}, a recall of \textbf{0.94}, and an F1-score of \textbf{0.95}. This balance indicates that {\tool} not only accurately detects a high proportion of actual runtime errors but also minimizes false positives. The overall accuracy of \textbf{97.51\%} shows superior capability in statically identifying runtime errors without execution.

Notably, the performance of all models remained relatively stable even when tested on incomplete code snippets (not shown). This indicates that the models, including {\tool}, can understand the semantic meaning of the removed library, generalize well, and maintain high detection accuracy.

We did not use CodeExecutor as a baseline for runtime error detection because it is only trained on datasets with full execution traces that lack instances of crashes and runtime errors. Consequently, it always provides the execution trace and intermediate values until the end of execution, rather than detecting or stopping at crash points, failing to detect errors.



\section{Runtime Error Localization Accuracy (RQ3)}
\label{sec:RQ3}
After detecting whether a snippet contains errors, the next step is to localize the specific lines that raise these errors.

\subsection{Data Processing and Evaluation Metrics}
We evaluated all models on approximately 1,300 different buggy code snippets from the FixEval dataset~\cite{haque2023fixeval}. In addition to the Complete Runtime Error Dataset in RQ2, we created the {\em Incomplete Runtime Error Dataset} by removing all import statements and external file references from each snippet. 


In this section, we focus on a new metric, \textbf{Error Localization Accuracy (ELA)}, which measures the number of times the {\em predicted buggy line matches the actual buggy line}. Based on the results from Section~\ref{sec:RQ2}, we observed that Claude performed the best in detecting runtime errors. Therefore, for this experiment, we use Claude as the main baseline to compare with our model.
 
\subsection{Empirical Results}

\begin{table}[ht]
\centering
\caption{Error Localization Accuracy Comparison (RQ3)}
\label{tab:BugLocalizationAcc}
\begin{tabular}{l c c c c}
\toprule
 \multirow{2}{*}{\textbf{Metric}} & \multicolumn{2}{c}{\textbf{Claude}} & \multicolumn{2}{c}{\textbf{\tool}} \\ 
\cmidrule(lr){2-3} \cmidrule(lr){4-5}
 & \textbf{Complete} & \textbf{Incomplete} & \textbf{Complete} & \textbf{Incomplete} \\ 
\midrule
ELA (\%) & 60.20 & 59.41 & \cellcolor{green!15}\textbf{72.22} & \cellcolor{green!15}\textbf{70.37} \\ 
\bottomrule
\end{tabular}
\end{table}

The results in Table~\ref{tab:BugLocalizationAcc} show that {\tool} significantly outperforms Claude in runtime error localization accuracy. For the Complete Runtime Error Dataset, {\tool} achieved an accuracy of \textbf{72.22\%}, compared to Claude's 60.20\%. This demonstrates {\tool}'s superior ability to accurately pinpoint the exact lines causing runtime errors for complete code.
Similarly, for the Incomplete Runtime Error Dataset, {\tool} maintained a high bug localization accuracy of \textbf{70.37\%}, while Claude's performance slightly dropped to 59.41\%. This consistency highlights {\tool}'s effectiveness even when external library imports are removed, indicating that the model can still understand and trace the flow in the code accurately.

\subsubsection{HeatMap Visualization} 
\begin{figure}[t]
    \centering
    \includegraphics[width=0.8\columnwidth]{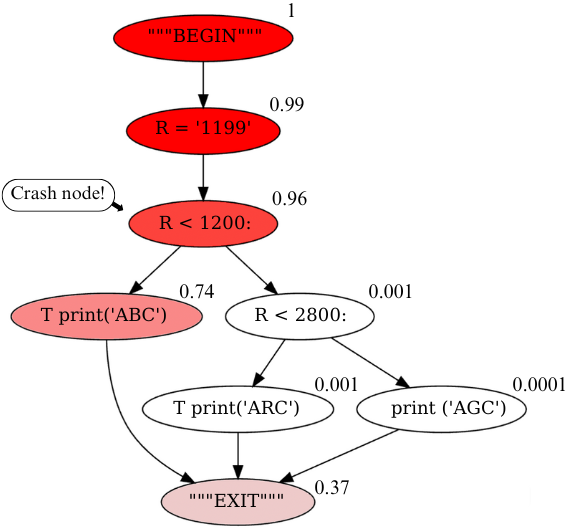}
    \caption{Heatmap visualization of node scores in buggy code.}
    \label{fig:runtime_error_visualization}
\end{figure}

In addition to Table~\ref{tab:BugLocalizationAcc}, further analysis of the heatmap visualization in Fig.~\ref{fig:runtime_error_visualization} for an example provides insights into {\tool}'s effectiveness in handling buggy code. In the heatmap, we observe that the {\em scores of nodes do not significantly drop in buggy code}, unlike in non-buggy code where the score for nodes not in the covered path drops very low (indicated by white color in Fig.~\ref{fig:visualization}d). In buggy code, the crash point reduces the score, but the nodes in the likely-correct path to the \textit{EXIT} node still retain high scores.

To further enhance error localization accuracy, we experimented with increasing the $\alpha$ value in Section~\ref{sssec:classification} to classify node. By filtering out more non-covered nodes, we observed improved accuracy as shown in Table~\ref{tab:BugLocalizationAlpha}.

\begin{table}[t]
\centering
\caption{Error Localization Accuracy with Different Alpha Values (RQ3)}
\label{tab:BugLocalizationAlpha}
\small
\begin{tabular}{l c}
\toprule
 \multicolumn{1}{c}{\textbf{Alpha Value}} & \multicolumn{1}{c}{\textbf{ELA (\%)}} \\ 
\midrule
$\alpha = 0.5$ & 72.22 \\
$\alpha = 0.7$ & 74.82\\
$\alpha = 0.9$ & 77.40\\
\textbf{$\alpha = 0.95$} & \textbf{78.31} \\ 
\bottomrule
\end{tabular}
\end{table}

As we increase the alpha value from 0.5 to 0.95, the bug localization accuracy improves, reaching 78.31\% at $\alpha = 0.95$. This indicates that by setting a higher threshold, {\tool} becomes more effective at filtering out non-relevant nodes, thereby enhancing its ability to identify the buggy lines. 

Additionally, to further demonstrate the effectiveness of {\tool} in runtime error detection, we have compiled a list of the top 10 runtime errors that our model can successfully identify. Table~\ref{tab:top10_error} highlights these common runtime errors along with their corresponding error messages. 
\begin{table}[t]
\centering
\caption{Top 10 Runtime Errors Detected by \tool}
\label{tab:top10_error}
\scriptsize
\begin{tabular}{>{\raggedright\arraybackslash}p{2.9cm} >{\raggedright\arraybackslash}p{5cm}}
\toprule
\rowcolor{gray!20} \textbf{Runtime Error} & \textbf{Error Message} \\
\midrule
Operand Type Mismatch & unsupported operand type(s) for ** or pow(): 'str' and 'int' \\
\addlinespace
Comparison Error & '<' not supported between instances of 'list' and 'int' \\
\addlinespace
Object Not Callable & 'int' object is not callable \\
\addlinespace
Non Iterable Type & 'int' object is not iterable \\
\addlinespace
Invalid Argument Type & list indices must be integers or slices, not str \\
\addlinespace
TypeError & 'float' object cannot be interpreted as an integer \\
\addlinespace
Type Specific Operation & can't multiply sequence by non-int of type 'str' \\
\addlinespace
Non Subscriptable & 'int' object is not subscriptable \\
\addlinespace
Attribute Error & object of type 'int' has no len() \\
\addlinespace
NoneType Subscripting & 'NoneType' object is not subscriptable \\
\bottomrule
\end{tabular}
\end{table}

\section{Usefulness in Fuzz Testing (RQ4)}
\label{sec:RQ4}
For this study, we evaluated the usefulness of {\tool} in supporting fuzz testing to detect and localize runtime errors in incomplete/non-executable code.

\subsection{Fuzz Testing Procedure}
The fuzz testing procedure consists of three main steps:

\begin{enumerate}[leftmargin=*]
    \item \textbf{Input Generation}: We used Claude to generate inputs that are likely to raise runtime errors in the provided snippets.
    \item \textbf{Runtime Error Detection}: The code snippet with the generated inputs was fed into {\tool}. The model processed the code to determine whether it contained a runtime error and, if so, localized the buggy statement.
    \item \textbf{Feedback Loop}: If no runtime error is detected in Step 2, the process enters a feedback loop. The inputs from Step 1 that failed to raise an error were fed back into the LLM to regenerate new inputs. This process continues until a runtime error is discovered or the time limit is exceeded.
\end{enumerate}

\subsection{Empirical Results}

\begin{table}[ht]
\centering
\caption{Runtime Error Detection Comparison (RQ4)}
\label{tab:RQ4results}
\begin{tabular}{lcccc}
\toprule
\multirow{2}{*}{\textbf{Metric}} & \multirow{2}{*}{\textbf{Claude}} & \multicolumn{3}{c}{\textbf{Fuzz Testing w/ \tool}} \\ 
\cmidrule(lr){3-5}
 & & \textbf{30s} & \textbf{60s} & \textbf{120s} \\ 
\midrule
\#runtime error detected & 32/50 & 44/50 & 46/50 & 47/50 \\ 
ELA (\%) & 42.27 & 49.53 & 50.00 & 42.61 \\ 
\bottomrule
\end{tabular}
\end{table}

To evaluate {\tool}'s effectiveness in supporting fuzz testing, we tested 50 buggy Python snippets from the FixEval dataset after removing all input variables and \code{import} statements. The results (Table~\ref{tab:RQ4results}) show that Claude alone detected {\bf 32} runtime errors, while its integration with {\tool} detected {\bf 44} errors in 30 seconds (a 37.5\% improvement), 46 errors in 60 seconds, and 47 errors in 120 seconds.

Additionally, the Error Localization Accuracy (ELA) with {\tool} was consistently higher than that of Claude, as showed in RQ3. It is particularly challenging for LLMs like Claude to detect the correct line containing bugs in incomplete code due to the lack of inputs, making it difficult for Claude to reason about code execution and runtime behaviors.

The significant enhancement from our model is especially valuable for incomplete code, where direct execution is infeasible. By combining Claude to generate inputs and {\tool} to predict runtime errors without external library setups, this approach effectively addresses the challenge. It not only improves runtime error detection rates but also provides accurate fault localization. LLM-based fuzzers, e.g., Fuzz4All~\cite{xia2024fuzz4all} could also seamlessly integrate in our~framework.


\section{Related Work}
\label{sec:related}

\subsubsection*{Predictive Execution}

CodeExecutor~\cite{liu2023code} was pre-trained on a dataset including source code, input values, and execution traces with values at each step. Its transformer learns to convert input and source code into execution traces. Ding et al.~\cite{ding2023traced} introduce TRACED, an execution-aware pre-training strategy using a mix of source code, executable inputs, and execution traces. We did not compare  with TRACED since it works only for C. LExecutor~\cite{souza2023lexecutor} predicts and injects missing values to execute arbitrary (in)complete code. It still requires execution. TraceFixer~\cite{bouzenia2023tracefixer} is trained using buggy code, execution traces, desired values, and expected bug-fixed code. Bieber {\em et al.}~\cite{bieber2020learning} learn to execute on CFGs. They introduce additional parameter complexity to learn soft dynamic branching, whereas we set branching conditions based only on the current hidden state. Tufano {\em et al.}~\cite{tufano2023predictingcodecoverageexecution} prompt to LLM to return the code coverage.

\subsubsection*{ML-based Fault Localization}

Early ML fault localization approaches~\cite{zheng2016fault, briand2007using, zhang2017deep, wong2009bp, TraPT} primarily rely on test coverage data and struggle to distinguish between elements executed by failed tests and actual faulty elements~\cite{TraPT}. In contrast, recent deep learning-based approaches such as GRACE~\cite{lou2021boosting}, DeepFL~\cite{DeepFL}, CNNFL~\cite{zhang2019cnn}, and DeepRL4FL~\cite{icse21-fl} have demonstrated improved performance. GRACE introduces a novel graph representation for methods and learns to rank faulty methods. 

Earlier learning-based FL techniques include MULTRIC~\cite{MULTRIC}, TrapT~\cite{TraPT}, and Fluccs~\cite{sohn2017fluccs}. Automated program repair approaches~\cite{icse22, hercules-icse19} focus on locating and fixing bugs. The Hercules APR tool~\cite{hercules-icse19} can identify multiple buggy hunks. FixLocator~\cite{fse22} detects co-fixing locations, and TRANSFER~\cite{meng2022improving} utilizes deep semantic features and transferred knowledge from open-source data to enhance FL. CodeT5-DLR~\cite{bui2022detect} introduces an end-to-end pipeline on LLMs to detect, localize and repair bugs in sequential order.

\section{Conclusion \& Future Work}

Current code models often overlook dynamic dependencies between lines of code, focusing only on plain text or correct execution traces. To address this, we introduce {\tool}, an approach that predicts code coverage and detects runtime errors by learning both static and dynamic dependencies. {\tool} uses CFGs and a GRU network to represent execution paths and learn vector embeddings for CFG nodes. It also leverages execution traces via CFG to capture dynamic dependencies among statements. Our evaluation shows that {\tool} significantly improves coverage prediction accuracy and runtime error localization, outperforming state-of-the-art models. Our data and code is available in our website~\cite{project-website}.

\section*{Acknowledgment}
Tien N. Nguyen was supported in part by the US National Science Foundation (NSF) grant CNS-2120386 and the National Security Agency (NSA) grant NCAE-C-002-2021.

\balance

\bibliographystyle{IEEEtran}

\bibliography{references,references-extra}

\end{document}